# Quasi-Periodic Optical Key-Enabled Hybrid Cryptography: Merging Diffractive Physics and Deep Learning for High-Dimensional Security

Haiqi Gao[1,3,4,†], Yu Shao[1,3,4,†], Jiaming Liang[1,4], Xuehui Wang[1,4], Junren Wen[1,3,4], Yuchuan Shao[1,3,4], Yueguang Zhang[2], Weidong Shen[2,*], and Chenying Yang[1,2,*]

[1] Hangzhou Institute for Advanced Study, University of Chinese Academy of Sciences, Hangzhou, 310024, China

[2] State key laboratory of Modern Optical Instrumentation, Department of Optical Engineering, Zhejiang University, Hangzhou, 310027, China

[3] Shanghai Institute of Optics and Fine Mechanics, Chinese Academy of Sciences, Shanghai, 201800, China

[4] Center of Materials Science and Optoelectronics Engineering, University of Chinese Academy of Sciences, Beijing, 100049, China

[†] These authors contributed equally to this work.
* E-mail: adongszju@hotmail.com (Weidong Shen); ycheny@zju.edu.cn (Chenying Yang)

## Abstract

Optical encryption inherently provides strong security advantages, with hybrid optoelectronic systems offering additional degrees of freedom by integrating optical and algorithmic domains. However, existing optical encryption schemes heavily rely on electronic computation, limiting overall efficiency, while the physical keys are susceptible to damage, compromising both security and system stability. To overcome these challenges, we introduce the Quasi-Periodic Optical Key (Q-POK), which combines long-range order with short-range disorder, enabling enhanced security and robustness against damage within a single platform. By leveraging diffraction symmetry, our design enables optics-driven encryption, effectively shifting the optoelectronic balance toward photonic processing. Moreover, we innovatively apply deep learning to reconstruct the complex optical ciphertext field using only amplitude data and cryptographic keys, simultaneously achieving data compression and improved security. Within this framework, the key space includes continuously tunable parameters such as wavelength, propagation distance, phase modulation, and Q-POK geometry, significantly expanding cryptographic diversity. Our system also demonstrates robust cryptographic reliability by reducing inter-class distances by over 50% and tolerating up to 20% ciphertext loss. Our framework represents a new generation of physically grounded, algorithmically enhanced optical cryptosystems—laying a foundational pathway for scalable, hardware-integrated information security paradigms.

# Introduction

Optical encryption utilizes the fundamental wave properties of light to achieve data protection at the physical layer. Compared to electronic approaches, this method not only provides inherent immunity to electromagnetic interference (EMI) but also effectively exploits the physical unclonability of the optical encryption layer for enhanced security.[1,2] Recent advances in metasurface technology have further improved system flexibility while expanding the encryption dimensionality through additional tunable optical degrees of freedom, including polarization, wavelength, phase, and angular momentum modulation[1,3–25]. Furthermore, hybrid optoelectronic architectures integrate these optical security advantages with nonlinear digital algorithms, enabling more robust cryptographic protection. However, current implementations face critical limitations - especially the over-reliance on electronic processing that compromises inherent optical advantages like high efficiency and low loss[6,8]. Meanwhile, the metasurface-based physical layer inevitably suffers from susceptibility to damage[7-9]. These challenges highlight the need for next-generation architectures that optimally balance optical and electronic components while addressing key security vulnerabilities in transmission and storage.

Optical diffraction, as an intrinsic wave property of light, naturally introduces distortion during the optical image propagation process, which provides inherent advantages for image encryption. The emergence of diffractive optical networks has further unlocked the potential of optical diffraction[26–34]. Multi-layer diffractive modulation has demonstrated its remarkable fitting capabilities, enabling its application in complex encryption processes[35,36]. Moreover, the conjugate property of optical diffraction ensures precise plaintext recovery when the ciphertext undergoes optical field conjugation[37]. However, optical conjugation requires complete light field recording (both amplitude and phase), significantly increasing information volume. This not only raises the difficulty of recording but also elevates the costs of storage and transmission. The key challenge lies in reconstructing the complex-conjugated optical field from amplitude-only measurements for decryption—a problem could be elegantly addressed by deep learning.

In this work, we propose a symmetric, lensless optoelectronic encryption and decryption framework that synergizes the strengths of physical-layer security and computational flexibility. The Quasi-Periodic Optical Key (Q-POK) serves as the cornerstone of our system, featuring a unique physical encryption structure that combines short-range randomness (via n×n stochastic phase modulations in each unit cell) with long-range periodicity (through m×m unit organization) and an intrinsic self-healing capability, enabling robust key recovery from partial damage when structural parameters are known.

We leverage the conjugate symmetry principle of optical diffraction during the decryption process, which is demonstrated in methods. Furthermore, we critically enhanced the process by a U-Net deep learning framework that explicitly incorporates the physical phase modulation of Q-POK as a prior constraint. Unlike traditional iterative approaches (e.g., Gerchberg-Saxton) that solely rely on amplitude matching under idealized propagation models, our approach achieves a direct mapping from the amplitude-plaintext and Q-POK pair ($|U|$, $\varphi_{Q\text{-}POK}$) to the missing phase $\varphi_{ciphertext}$ by jointly optimizing both physical consistency (through the diffraction model) and statistical patterns from experimental data, as shown in Fig.1 (a). This approach not only streamlines the decryption process but also reinforces the pivotal role of the Q-POK in optical security, which significantly enhances the efficiency of both encryption and decryption processes. Furthermore, simulation experiments demonstrate that the trained network can successfully perform cross-dataset decryption, trained on Digital-MNIST while evaluated on E-MNIST, as illustrated in Fig. 1(b). By incorporating diffraction distances ($d$), wavelength ($\lambda$), Q-POK geometry ($m \times n$), and phase modulation ($\varphi$) into the key space, our architecture achieves an unprecedented security dimensionality($m \times n \times d \times \lambda \times \varphi$) while maintaining the physical interpretability of the encryption process. This paradigm not only enhances security through spatial-light interdependence but also establishes a new benchmark for robust, diffraction-driven cryptographic systems.

## Results

**Framework of Q-POK-based Hybrid Cryptography**

The proposed encryption and decryption technology can be divided into two parts: the optical encryption flow and the optical decryption flow, as specifically illustrated in Fig. 1(a). In the optical encryption flow, the complex optical field of plaintext is transmitted through the Q-POK (denoted as $K$) by diffraction and encrypted into the complex optical field of ciphertext, according to the following expression:

$$C = P * H(\lambda, d) \otimes K$$

Here, * denotes the convolution operation, $\otimes$ denotes Hadamard product, and $H$ represents the Optical Transfer Function (OTF) of the system which depends on the operation wavelength $\lambda$ and the diffraction distance $d$. $C$ and $P$ is the complex field of ciphertext and plaintext, respectively. Due to the random selection of working wavelength, diffraction distance as well as Q-POK modulation, the encryption system achieves a large key space and thus offers high encryption capacity. All these parameters together form the encryption key, the influence of the operation wavelength $\lambda$ and the diffraction distance $d$ is further explored in supporting material S1, where the influences of ciphertext noise and alignment error on the system are also verified.

In the optical decryption flow, a U-Net model is employed to recover the phase information of the ciphertext. The U-Net receives the amplitude of the ciphertext and all the Q-POK modulation parameters, and it outputs the phase of the ciphertext required for the decryption process. Thus, we finally get the complex-conjugated optical field of ciphertext through combining the amplitude recorded by the camera and the phase generated by the U-Net. This U-Net architecture is composed of an encoder-decoder framework with symmetric skip connections, as shown in Fig. 1(c). The encoder path is composed of consecutive blocks, each consisting of two 3×3 convolutional layers, followed by batch normalization and ReLU activation functions, with 2×2 max pooling for down sampling. The decoder path uses transposed convolutions for up sampling, mirrored by corresponding skip connections to preserve spatial details. Finally, the U-Net output and the amplitude together compose the ciphertext*.

By conjugating the ciphertext and reversing the encryption operations (applying K in sequence), the plaintext $P$ can be recovered as:

$$P = C^* * H(\lambda, d) \otimes K,$$

which is similar to the encryption procedure. It should be noted that the Q-POK is used twice: during the U-Net-based phase evaluation, and the diffraction-based modulation. This dual application of the Q-POK significantly enhances the security of the system by reinforcing the decryption process at multiple stages. Furthermore, simulation results demonstrate that our decryption system is capable of recovering ciphertext information even when the ciphertext is randomly corrupted or partially lost, as illustrated in supporting material S2.

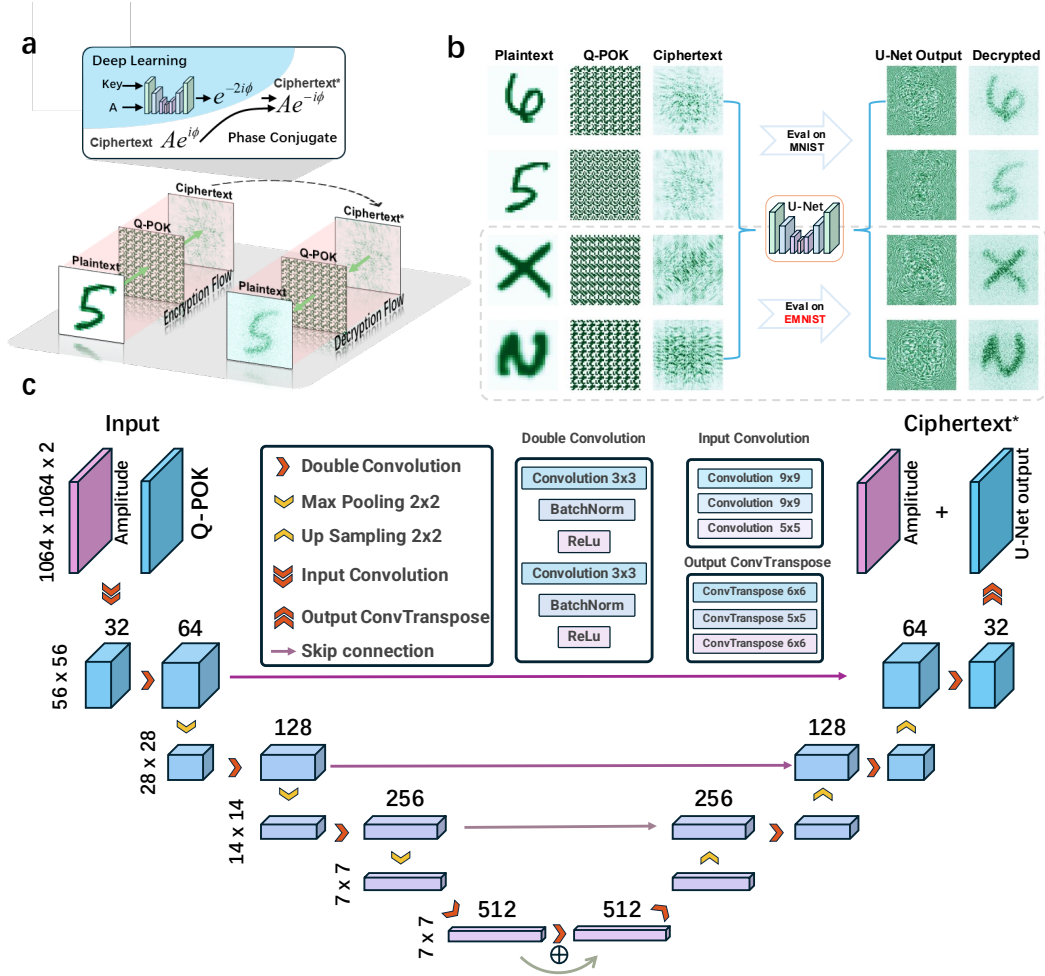

Figure 1 Framework of Q-POK-based Hybrid Cryptography. (a) The encryption and decryption flow processes. **Encryption flow**: the complex optical field of plaintext is modulated by Q-POK. **Decryption flow**: the amplitude of the ciphertext and the Q-POK are first sent to the U-Net, which predicts the phase of the ciphertext. The ciphertext is then transformed into its complex conjugate form. Following phase conjugation, it is modulated by the Q-POK through a diffraction-driven process. (b) The decryption results on valid dataset of Digital-MNIST (first two rows) and E-MNIST (last two rows). The U-Net used in both situations was trained with Digital-MNIST dataset. (c) Structure of the U-Net used for decryption. The input convolutional layers encode the amplitude information of the ciphertext along with the Q-POK modulation parameters, while the transposed convolutional layers progressively reconstruct the phase information as the output. Skip connections are incorporated to preserve spatial features across network layers.

**Multi-Dimensional Optical Key Design**

The proposed encryption and decryption architecture is fundamentally governed by optical diffraction effects, where any physical parameters influencing diffraction can serve as potential dimensions for key space expansion. Within our system, the diffraction propagation distance($d$), incident light wavelength($\lambda$), and phase modulation profile($\varphi$) emerge as three critical cryptographic parameters that collectively determine both the encryption results and decryption performance. These diffraction-dependent variables exhibit strong coupling effects that significantly enhance the security of the system while maintaining the physical interpretability of the encryption process.

Furthermore, we propose an innovative physical cryptographic layer design, as illustrated in Fig. 2a, to expand the dimension of key space. The Q-POK is designed based on a methodology combining long-range periodicity with short-range randomness. Initially, a key parameter n is selected to generate an n×n random phase plate as the fundamental unit. Subsequently, another critical parameter m is chosen to replicate and extend this unit into an m×m array, ultimately forming the physical optical layer with final dimensions of (m×n) × (m×n). This hierarchical architecture creates a bifunctional structure that simultaneously maintains local stochastic characteristics at the unit level while exhibiting global periodicity across the entire optical layer.

As the physical optical layer, Q-POK can be fabricated using the metasurface depicted in Fig. 2b. We employ $TiO_2$ as the structural material due to its high refractive index and low absorption in the visible spectrum, while fused silica is served as the substrate. Under illumination with a 671 nm linearly polarized light source, using structures with a 600 nm period and 1 μm height, we obtained the corresponding phase and transmission distributions. The results demonstrate that the spatial phase modulation range of the metasurface spans from $-\pi$ to $\pi$ (a full $2\pi$ coverage) while maintaining high transmittance. We selected two distinct structures capable of approximating 0 and $\pi$ phase modulation, respectively. The cross-sectional electric field profiles of these structures were simulated (Fig. S4), confirming their low crosstalk characteristics. And the metasurface-based encryption-decryption results were presented in Fig. S5, comfirming the feasibility of the selected metasurface structures for this Q-POK-based hybrid cryptography.

Fundamentally, the design strategy that integrates long-range periodicity with short-range randomness also functions as an intrinsic form of data compression. This hierarchical structure grants the Q-POK exceptional recoverability under physical damage. For random destruction (e.g., scratching), the key can be reconstructed by statistically analyzing the residual information from the same position across m × m repeating units, enabling rapid recovery of the original key, as illustrated in Fig. 2c. For physical destruction (e.g., fractures), the key can still be fully recovered as long as any single complete unit of the periodic structure is preserved, as demonstrated in Fig. 2d. Notably, the recovery process requires both the damaged Q-POK and key parameters (m, n), without knowledge of the original state of each unit. This recoverability feature maintains cryptographic security while highlighting the essential role of the physical optical layer.

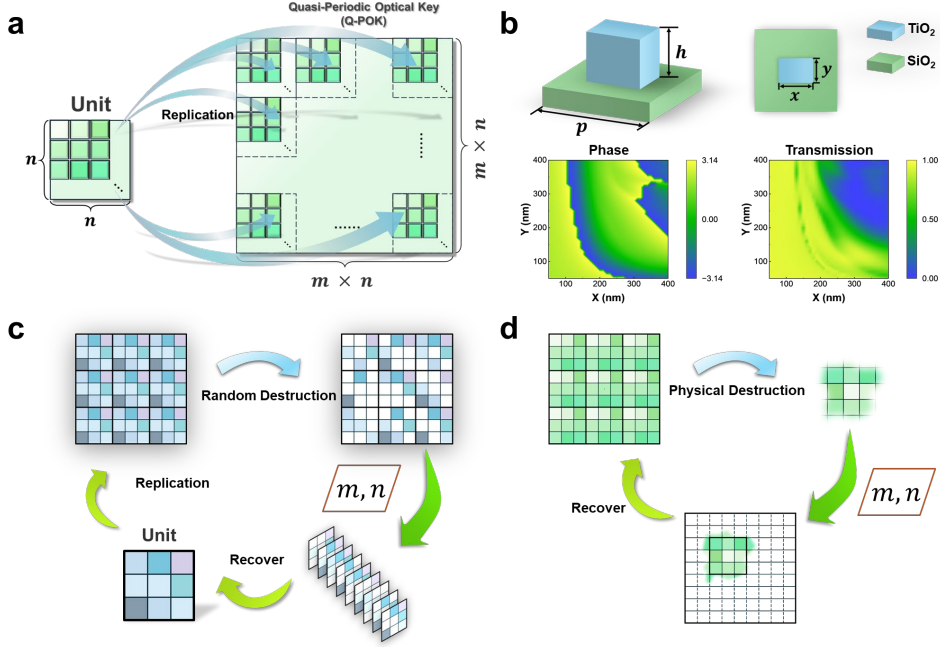

Figure 2 Design and Recovery Mechanism of the Quasi-Periodic Optical Key (Q-POK). (a) Design principle: A fundamental unit with size n×n is generated as a random phase plate and replicated m×m times to construct a quasi-periodic optical key with final dimensions of (m×n) × (m×n). This hierarchical arrangement maintains local randomness while establishing global periodicity across the optical layer. (b) Schematic illustration of the metasurface structure for Q-POK and simulated results of phase modulation and transmittance for varying unit cell sizes (p=600nm, h = 1000 nm, x ∈ [50, 400] nm, y ∈ [50, 400] nm). (c) Recovery under random partial damage: in cases of random partial damage, the original key can be statistically reconstructed by analyzing corresponding elements across multiple replicated units, leveraging the periodic redundancy introduced by the m×m replication process. (d) Recovery under physical destruction: even under severe physical damage where partial region of the optical layer is missing, full key recovery is achievable as long as any intact unit remains. The unit reset process utilizes known parameters n and m to accurately reconstruct the complete Q-POK without requiring prior knowledge of the original layout.

## Encryption Security Analysis

To further validate the uniqueness and cryptographic robustness of the Q-POK, we conducted digital experiments as illustrated in Fig. 3. Two independent networks were trained using distinct Q-POK parameter sets (n=7, m=8) and (n=8, m=7). Throughout this verification process, uniform Q-POK geometries were maintained with phase modulation strictly limited to binary states (0 and π). These two sets were then cross-tested by applying the mismatched Q-POK sets during the decryption stage. Notably, even with complete knowledge of all other system parameters, including diffraction distances ($d$) and wavelength ($\lambda$), successful decryption was achieved only when the Q-POK configuration precisely identical to the one used during encryption was applied, as shown in Fig. 3 a-d. This observation strongly confirms the robustness and security of the proposed Q-POK design. In contrast to previous optical encryption works[11,13,24], our encryption system is capable of encoding and decoding using randomly generated units without the need for retraining on each individual key. We trained a U-Net with a Q-POK parameter set (n = 7, m = 8), and subsequently encrypted and decrypted the same input pattern using different Q-POK configurations. Remarkably, the U-Net still accurately reconstructed the result, as shown in Fig. 3(e). This feature significantly reduces training resource consumption, demonstrating the superiority of our encryption-decryption architecture.

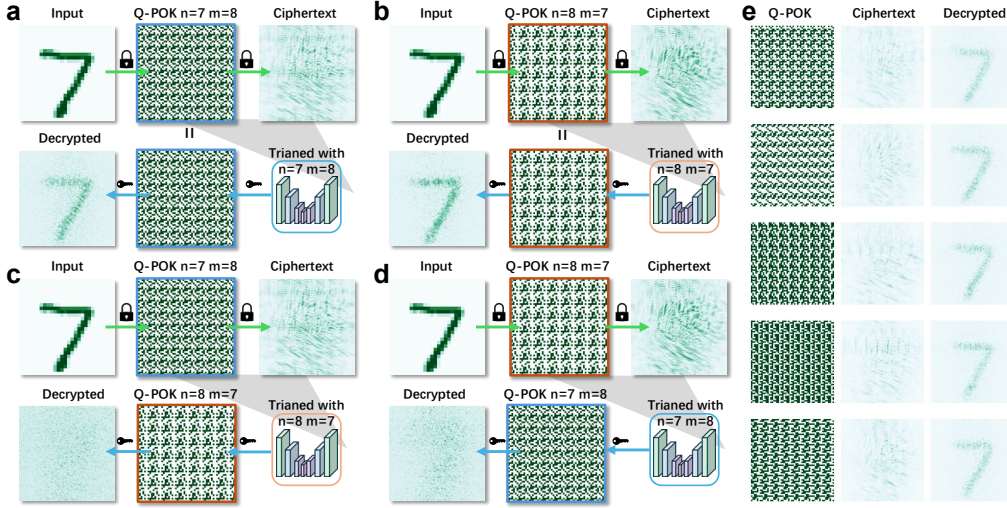

Figure 3 Validation of the Q-POK cryptographic security. (a) Encryption and successful decryption using a Q-POK with parameters n=7, m=8. (b) Encryption and successful decryption using a Q-POK with parameters n=8, m=7. (c) Failed decryption when encrypting with a Q-POK of n=7, m=8 and decrypting with a mismatched Q-POK of n=8, m=7. (d) Failed decryption when encrypting with a Q-POK of n=8, m=7 and decrypting with a mismatched Q-POK of n=7, m=8. Successful decryption occurs only when the original encryption Q-POK is used, demonstrating the high cryptographic security of the proposed structure. (e) The encryption-decryption results for identical input with different Q-POK using the same decryption U-net.

To quantitatively assess the encryption performance of our diffraction-based optical method, we performed statistical analysis including information entropy, histogram uniformity, cross-correlation, and principal component analysis (PCA) across varying propagation distances ($d_1 = 1400$ mm, $d_2 = 2800$ mm, and $d_3 = 4200$ mm), with those encryption-decryption results shown in Fig. S6-7. These metrics evaluate the statistical randomness, pixel decorrelation, and feature dispersion introduced by Q-POK encryption.

Specifically, entropy analysis demonstrates a significant increase in information entropy for the ciphertexts compared to the plaintext, as shown in Fig. 4a. The mean entropy of the plaintext image is 1.53, while the encrypted images at distances $d_1$, $d_2$, and $d_3$ yield entropy values of 5.97, 6.03, and 6.03, respectively. The improvement suggests that longer diffraction distances resulted in slightly better performance, as the increased distance enhanced spatial scrambling. Histogram variance is computed to assess the uniformity of intensity distributions, as shown in Fig. 4b. The plaintext exhibits a high mean histogram variance of approximately $3.39 \times 10^9$, reflecting a non-uniform pixel intensity distribution. Conversely, the encrypted images demonstrate significantly reduced variances—$1.18 \times 10^8$, $9.30 \times 10^7$, and $9.51 \times 10^7$ for $d_1$, $d_2$, and $d_3$, respectively, implying effective histogram flattening and therefore reduced statistical predictability. Cross-correlation analysis between the plaintext and each ciphertext reveals near-zero correlation as the propagation distance increases, dropping from 0.220 for $d_1$ to -0.004 for $d_3$, as shown in Fig. 4c. This decline reflects an effective decorrelation, indicating that ciphertexts are increasingly unrecognizable with respect to the original input, thus improving resistance to known-plaintext attacks.

Principal component analysis is also launched in our study, as shown in Fig.4 d. Compared with the plaintext, the ciphertext's PCA projection shows greater dispersion in the feature space, indicating more effective spatial scrambling. It is noteworthy that the PCA projection scatter plot reveals a more compact clustering of encrypted data samples within the first two principal components, exhibiting lower dispersion and a more centralized feature distribution. Additionally, the average Euclidean distance between samples significantly decreases from over 38,000 in the plaintext data to approximately 13,000 after encryption, further demonstrating the tighter clustering effect. This phenomenon suggests that the encryption process, to some extent, suppresses data diversity and reduces structural differences among data samples. The reduced dimensionality representation suggests the encrypted data has a more homogeneous and less structured distribution, further supporting effective obfuscation of salient image features.

Generally, the statistical indicators consistently demonstrate that longer encryption distances result in ciphertexts with higher entropy, lower histogram variance, minimal correlation with plaintext, and reduced PCA-represented variance, all of which contribute to a stronger security profile.

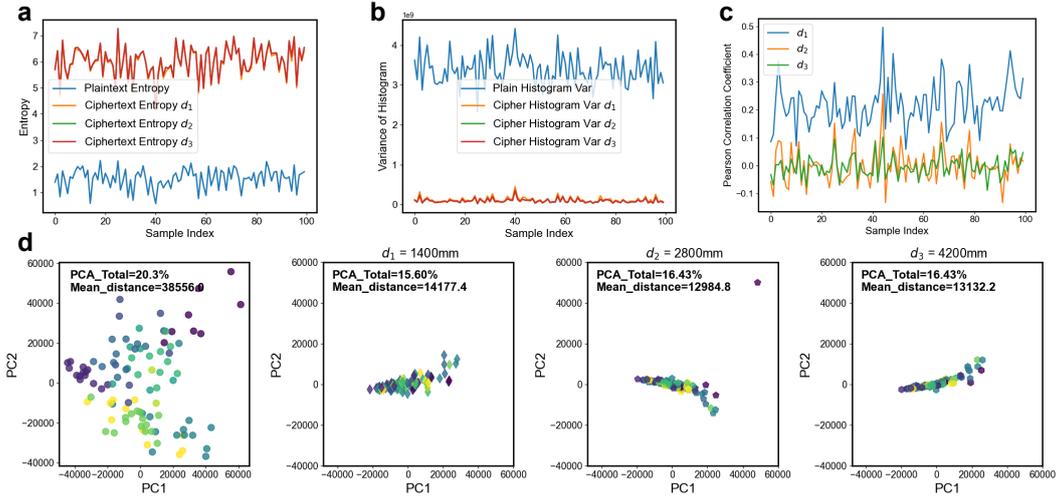

Fig. 4 Statistical analysis of encryption security. (a) Information entropies of different samples, higher information entropies indicate higher randomness. (b) Histogram variance analysis of different samples, lower variance corresponds to more uniform intensity distributions. (c) Cross-correlation coefficients between plaintext and ciphertexts of different samples, lower cross-correlation coefficients reflect improved decorrelation. (d) PCA analysis of different samples, the selected samples consist of ten categories, with each color representing one category.

**Experimental validation**

To evaluate the feasibility of our proposed encryption and decryption architecture, we conducted a proof-of-concept experimental demonstration. For simplicity, a single-layer Q-POK with phase values restricted to 0 and $\pi$ was employed, with key parameters set to m=8 and n=7. As shown in Fig.5a, we used 3D-printed digital masks as input, loaded the Q-POK onto a spatial light modulator (SLM), and obtained the encrypted patterns after light propagation, with representative examples shown in the third column of Fig. 5b. To simplify the decryption verification, we modified the optical setup by directly placing a phase-type SLM at the output of the encryption path and loaded the output of U-Net into SLM with twice the negative phase of the ciphertext, enabling the decryption process to directly use the conjugate of the ciphertext as input. The experimental light path is shown in Fig. S8. This configuration maintains strict symmetry with the encryption path, ultimately restoring the plaintext.

Two different methods were employed for decryption: theoretical complex-field reconstruction through ideal phase conjugation and U-Net-based complex-field recovery. The ideal phase conjugation method involved theoretically calculating the complex optical field of the ciphertext, which was subsequently used to derive the SLM parameters for decryption. This approach resulted in the decryption path receiving the exact conjugate of the complex field of the ciphertext, with corresponding decryption results presented in column 4 and 5 of Fig. 5b, for experimental and simulated results, respectively. In contrast, the U-Net approach eliminated the need for prior knowledge of the complex field of the ciphertext; by simply feeding the amplitude of the ciphertext and Q-POK into the trained U-Net network, we obtained SLM parameters that produced an optical field closely approximating the conjugate of the complex field of the ciphertext. The decryption outcomes of the U-Net are shown in column 6 and 7 of Fig. 5b for experimental and simulated results, respectively. Notably, both methods shared the same optical configuration while employing distinct computational strategies for phase retrieval and decryption, but they produced statistically consistent correct results, demonstrating the capability of the U-Net in retrieving the complex conjugate of the optical field of the ciphertext.

Throughout the encryption and decryption process, high alignment precision of the optical path was required. To compensate for scaling effects caused by imperfect collimation, we introduced a Fresnel phase correction by applying an additional phase profile to the SLM. This correction moderately improved our experimental results, as demonstrated in Fig. S9. In theory, an ideal optical system would produce images that perfectly match the simulation results. However, practical limitations, including prism deflection, collimation inaccuracies, and alignment errors, led to only approximate pattern reconstruction. Besides, the influences of wavelength $\lambda$, diffraction distance $d$, noise on ciphertext $\sigma$, and alignment error were also explored, as shown in Fig. S9. Despite the aforementioned errors, the decrypted results remain clearly distinguishable via an industrial camera, as demonstrated in Fig. 5b, confirming the reliability of the decryption process.

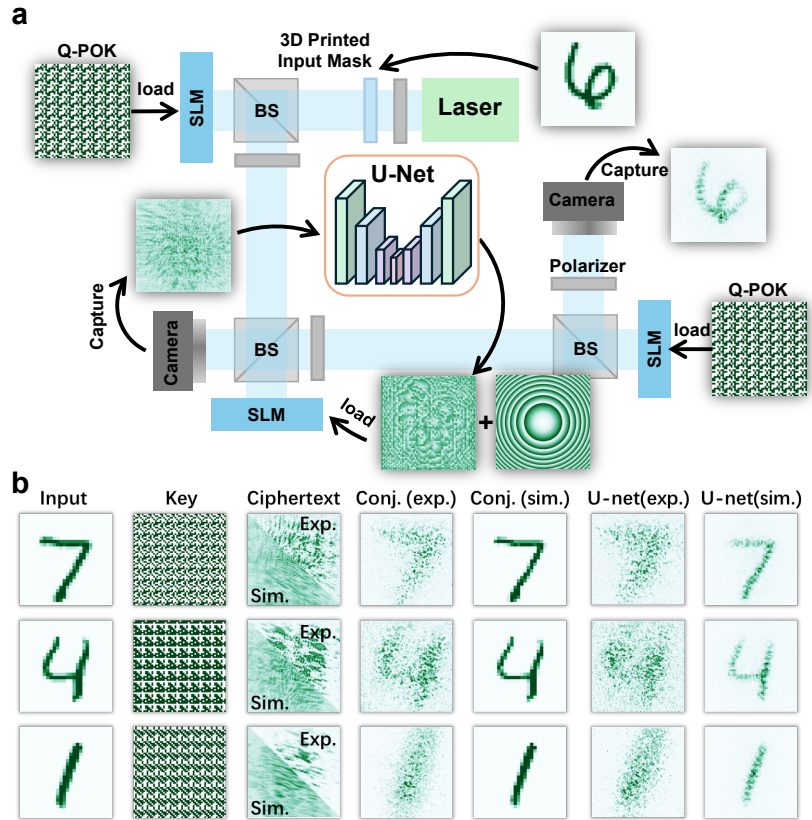

Figure 5 Experimental validation of the proposed Q-POK-based Hybrid Cryptography. (a) Schematic diagram of experimental setup, (b) Experimental and simulated encryption/decryption results, with the conjugated phase derived from theoretical calculation and U-net-based recovery.

## Discussion

This work presents a novel encryption-decryption system. Compared to conventional double random phase encoding (DRPE), our system features a simpler optical configuration while integrating electronic networks to address the inherent linearity limitation of optical systems, thereby enhancing operational flexibility. Compared to recent multi-dimensional optical encryption techniques, our solution overcomes the structural fragility of physical optical layer and better leverages inherent optical properties, particularly the diffraction symmetry. Additionally, owing to the high-dimensional key space, we could achieve a one-to-one correspondence between keys and patterns, which would significantly enhance the security of the encryption system. Though a spatial light modulator (SLM) was used as the physical optical layer for validation, the Q-POK is intended to be implemented using metasurface-based hardware, as proposed (Fig. S4). Alternatively, it can be effectively replaced by relief structures (Fig. S10), as demonstrated in our previous work[38].

Furthermore, the designed Q-POK exhibits inherent compatibility with multiple optical dimensions. While our current implementation only combines it with fundamental diffraction parameters (propagation distance $d$ and wavelength $\lambda$), the Q-POK can potentially integrate with other intrinsic optical properties including polarization, incident angle, and orbital angular momentum to further expand the key space dimensionality. It should be emphasized that the Q-POK is fundamentally a compressible cryptographic key, which provides exceptional damage resilience through its recoverability feature. However, this recoverability requires the damaged key to retain at least one complete minimum unit (n×n phase pattern), imposing the constraint that n must be greater than 1 (n>1) for successful reconstruction. Besides, the Q-POK could also be designed as a multi-layer structure to introduce additional complexity into the architecture.

In conclusion, we propose a novel optoelectronic encryption and decryption system. This framework leverages the physically reversible nature of optical diffraction and its phase conjugation characteristics, thereby emphasizing the role of diffraction in the encryption process. As well, a distinctive Quasi-Periodic Optical Key (Q-POK) is introduced to extend the functional capabilities of both encryption and decryption. Additionally, a U-Net is employed to reconstruct the complex conjugate optical field, a process that not only enhances cryptographic security but also reduces the complexity of data acquisition and storage. These innovations offer new strategies and insights for developing integrated optical and optoelectronic encryption systems.

## Methods

### Time-Reversal Symmetry in Wave Propagation

Each linear optical effect, including diffraction, phase modulation, and amplitude modulation, can be considered as a linear operator $T_i$. The forward propagation of an optical field through a series of such behaviors can be expressed as

$$E_{out} = T_n \cdots T_2 T_1 E_0$$

When the output field is subjected to phase conjugation (i.e., complex conjugation) and then propagated backward through the same systems in reverse order, the resulting field becomes:

$$E'_{out} = T_1 T_2 \cdots T_n (T_n \cdots T_1 E_0)^*$$

Under the ideal assumption that each operator satisfies the unitary-like condition $T_i^* T_i = I$, the

expression simplifies to:

$$E'_{out} = E_0^*$$

**Training details of the U-net for ciphertext* restruction**

To recover the phase information from the ciphertext after Q-POK encryption, a U-Net is trained in this study. The target of the U-Net is $-\varphi$, where the $\varphi$ is the phase term of the optical field of ciphertext. During the training process, we employ a periodic mean squared error (PMSE) loss function, defined as $\mathcal{L}_{PMSE}(\mathbf{y}, \hat{\mathbf{y}}) = \frac{1}{n}\sum_{i=1}^{n}\left[\left((\hat{y}_i - y_i + \pi)\mod 2\pi - \pi\right)^2\right]$, to optimize model performance. The proposed periodic mean squared error (PMSE) loss function addresses the inherent challenges of phase optimization by incorporating cyclic boundary conditions in a differentiable manner. To minimizing the loss function, parameters of the U-Net are optimized via Adam optimizer, whose learning rate is 0.001 and decays with cosine annealing for 20 epochs. The training framework is Pytorch 2.6.0 with Python 3.12, using CUDA for GPU acceleration.

**Experimental setup**

In the optical experiment, we used three spatial light modulators (SLMs), each with a resolution of 1920 × 1080 and a pixel size of 8 μm. The first and third SLMs were loaded with Q-POK, while the second SLM was loaded with the calculation result from the U-Net. The laser was positioned 400 mm from the first SLM. The distance between the first and second SLMs was 1000 mm, as was the separation between the second and third SLMs. Finally, the third SLM was placed 400 mm from the recording camera. The input pattern was resized to 56 × 56, and each pattern pixel consisted of 19 × 19 physical pixels, resulting in a total of 1064 × 1064 pixels used per SLM.